\documentclass[draft,jgrga]{agutex2015}
\usepackage[dvips]{graphicx}
\usepackage{amssymb,amsmath}
\usepackage{url}
\usepackage{xcolor}
  \setkeys{Gin}{draft=false}

\authorrunninghead{KATUSHKINA ET AL.}
\titlerunninghead{VOYAGER~1 Lyman~$\alpha$ MEASUREMENTS }

\begin{document}

\title{Voyager~1/UVS Lyman~$\alpha$ measurements at the distant heliosphere (90-130~AU): unknown source of additional emission}
\authors{O. A. Katushkina,\altaffilmark{1}
 E. Qu\'{e}merais,\altaffilmark{2} V. V. Izmodenov,\altaffilmark{1,3,4}
 R. Lallement\altaffilmark{5} and B. R. Sandel\altaffilmark{6}}

\altaffiltext{1}{Space Research Institute of Russian Academy of Sciences, Moscow, Russia.}

\altaffiltext{2}{Universit\'{e} Versailles Saint-Quentin, LATMOS, Guyancourt, France.}

\altaffiltext{3}{Department of Mechanics and Mathematics, Lomonosov Moscow State University, Moscow, Russia.}

\altaffiltext{4}{Institute for Problems in Mechanics, Moscow, Russia.}

\altaffiltext{5}{GEPI/Observatoire de Paris, Meudon, France.}

\altaffiltext{6}{Lunar and Planetary Laboratory, The University of Arizona, Tucson, Arizona, USA.}

\keypoints{\item For the first time we present the Lyman~$\alpha$ intensities measured by Voyager~1/UVS in 2003-2014 (at 90-130~AU from the Sun).
 \item The data show an unexpected flat behavior at 90-115~AU. \item An additional source of emission is necessary to fit the data.
 We suggest two possible qualitative scenarios that can explain the data.   }

\begin{abstract}
In this work, we present for the first time the Lyman~$\alpha$ intensities measured by Voyager~1/UVS in 2003-2014 (at 90-130~AU from the Sun).
During this period Voyager~1 measured the Lyman~$\alpha$ emission in the outer heliosphere at an almost fixed direction close to the upwind (that is towards
the interstellar flow).
The data show an unexpected behavior in 2003-2009: the ratio of observed intensity to the solar Lyman~$\alpha$ flux is almost constant.
Numerical modelling of these data is performed in the frame of a state-of-art self-consistent kinetic-MHD model of the
heliospheric interface (Izmodenov \& Alexashov, 2015). The model results, for various interstellar parameters, predict a monotonic
decrease of intensity not seen in the data. We propose two possible scenarios that explain the data qualitatively.
The first is the formation of a dense layer of hydrogen atoms near the heliopause. Such a layer would provide an additional backscattered Doppler shifted Lyman~$\alpha$ emission, which is not absorbed
 inside the heliosphere and may be observed by Voyager. About 35~R of intensity from the layer is needed.
The second scenario is an external non-heliospheric Lyman~$\alpha$ component, which could be galactic or extragalactic. Our parametric study
shows that $\sim$25~R of additional emission leads to a good qualitative agreement between the Voyager~1 data and the model results.
\end{abstract}

\begin{article}

\section{Introduction}

\begin{figure}
\noindent\includegraphics[scale=0.8]{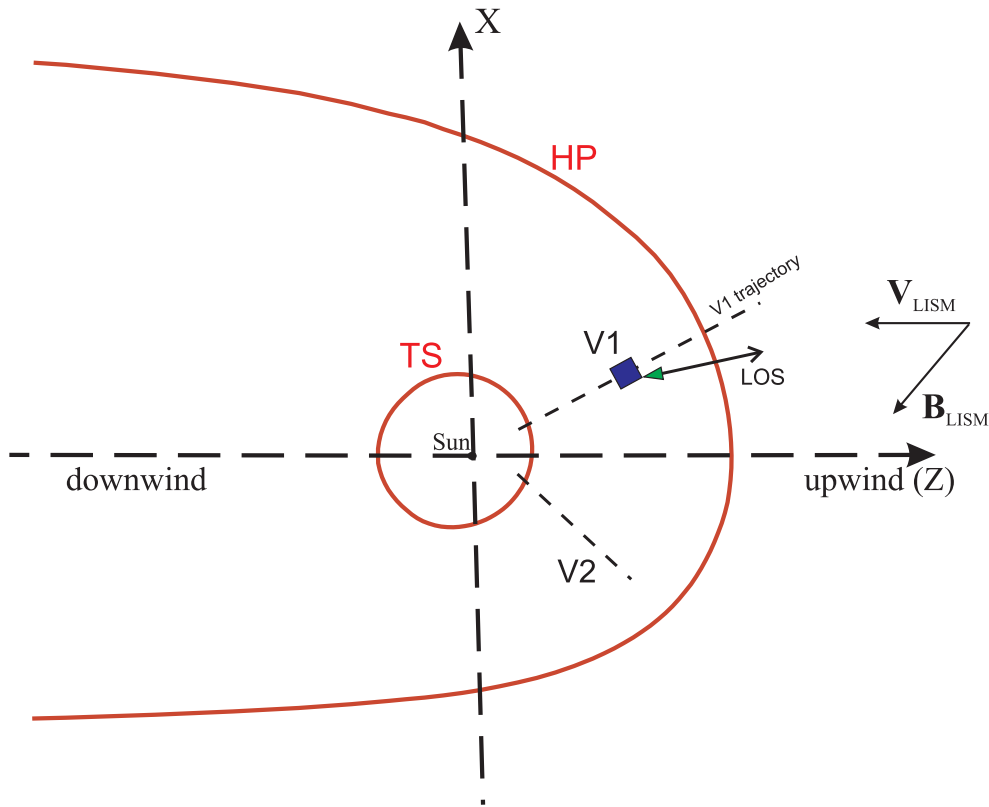}
\caption{Schematic illustration of the SW/LISM interaction region (not scaled). The termination shock (TS) and the heliopause (HP) are marked.
Projections of Voyagers (V1 and V2) trajectories on the $(\textbf{B}_{LISM},\textbf{V}_{LISM})$-plane are shown.
Line of sight of Voyager~1/UVS ia also shown.}
\label{interface}
\end{figure}

The Voyager mission is one of the most outstanding projects devoted to investigating the Solar System and the distant heliosphere.
The twin Voyager~1 and 2 spacecraft were launched in 1977, and now, after 39 years of operating, they are the furthest human-made
objects in the universe. The original objective of the Voyagers was to explore the outer planets of the Solar System (Jupiter, Saturn, Uranus, and Neptune), but afterwards
the mission was extended and Voyagers began their interstellar journey.
Following their last gravitational flyby, both Voyager~1 and 2 move radially away from the Sun in the nose part of the heliosphere above and below
the ecliptic plane, correspondingly (see Fig.~\ref{interface}). Their trajectories are inclined to the upwind direction (that is opposite to
 the interstellar flow) at 20-30$^{\circ}$. Voyager~1's velocity is about 3.6 AU per year, while Voyager 2's velocity is about 3.3 AU per year.
 During their long trip Voyagers study the region of interaction between the solar wind (SW) and the local interstellar medium (LISM),
which is called the heliospheric interface (see Fig.~\ref{interface}).
Voyager~1 and Voyager~2 crossed the first boundary of the heliospheric interface - the termination shock (TS) - in 2004 at 94~AU from
 the Sun and in 2007 at 84~AU from the Sun, respectively.
It is supposed \citep{stone_etal_2013, gurnett_etal_2013, burlaga_ness_2014_apjl} that in 2012 Voyager~1 crossed the
second boundary, that is, the heliopause
(HP, tangential discontinuity separating the solar wind plasma from the charged component of the interstellar plasma) and entered the interstellar medium.
Voyager~2 is still in the inner heliosheath region (between TS and HP) and approaching the heliopause \citep{richardson_decker_2014, burlaga_etal_2016}.

The Voyagers provide unique information on the heliospheric boundary owing to their direct measurements of plasma parameters
(available now only on Voyager~2), energetic particles and magnetic field.
Besides this, both Voyagers are equipped with the Ultraviolet Spectrograph (UVS) that is described in detail by \citet{broadfoot_etal_1977}.
The UVS is an objective grating spectrograph covering the wavelength range 540-1700~${\rm \AA}$. A single input photo-electron is converted to a pulse
of many electrons by microchannel plates (MCPs). The output charge is collected in 128 elongated channels. The Lyman~$\alpha$ signal (wavelength at line center: 1215.6~${\rm \AA}$) is measured through integration over the nine spectral channels defining the response to a monochromatic narrow emission line at the Lyman~$\alpha$ wavelength.

The main source of the heliospheric Lyman~$\alpha$ emission is scattering of solar photons by the interstellar hydrogen (ISH) atoms, which penetrate into the
heliosphere from the interstellar medium due to relative motion of the Sun and the LISM. Measurement of Lyman~$\alpha$ emission
at the outer heliosphere is a tool for remote sensing of the ISH distribution, which in turn is an important source of information on the LISM parameters and
the physical processes at the heliospheric boundary.

The UVS instrument on Voyager~2 was switched off in 1998, while the UVS on Voyager~1 has operated from 1977 until 2016. In October, 2014, a sudden sharp decrease of the signal was detected. Most probably this was caused by technical troubles with equipment and the data obtained after that
 are more difficult to interpret and so not considered here. Voyager~1/UVS collected a huge amount of data during 37 years. Observations can be divided into three periods:
1) 1979-1993 ($\sim$7-53~AU), observations for a variety of lines-of-sight with a few rolls over the sky \citep{lallement_etal_1991, hall_etal_1993, quemerais_etal_2013, fayock_etal_2015}; 2) 1993-2003 ($\sim$54-89~AU), systematic scans over the sky from the upwind to downwind directions \citep[e.g.,][]{quemerais_etal_2010, lallement_etal_2011, katushkina_etal_2016}; 3) 2003-2014 ($\sim$90-130~AU), Voyager~1's scan platform
was not moved and all measurements were performed for approximately the same line-of-sight (LOS) close to the upwind direction.

In this work for the first time we consider the last period of measurements, which covers more than 4000 observations of Lyman~$\alpha$
intensity during 11 years. These data
have not been published and analyzed before. This period of measurements is very interesting because it contains the data obtained close to the heliopause
and it could give unique information about both the ISH distribution at the heliospheric boundary and a possible minor component of the extra-heliospheric Lyman~$\alpha$ emission that can be detected only far enough from the Sun \citep{lallement_etal_2011}.

We also performed modelling of Lyman~$\alpha$ intensities measured during New Horizons cruise observations by the Alice spectrograph at 11 and 17 AU from the Sun.
These data confirm our main conclusions derived from the Voyager~1 data analysis.

The paper is organized as follows. In Section~2 the Voyager~1/UVS data obtained in 2003-2014 are presented and described in detail.
Section~\ref{model} provides a brief description of the numerical model for the IPH distribution and the radiative transfer
that is applied for the data analysis. Comparison of the model results and observations is presented in Section~\ref{comparison}.
In Section~\ref{scenario} we propose two possible scenarios, which are consistent with the data. Section~\ref{alice} shows
the New Horizons/Alice data (obtained during cruise observations) and comparison with the model results. Discussion and summary are presented in Section~\ref{discussion}.

\section{Voyager~1/UVS observations in 2003-2014}
\label{data}

\subsection{Observational geometry}

\begin{figure}
\noindent\includegraphics[scale=0.8]{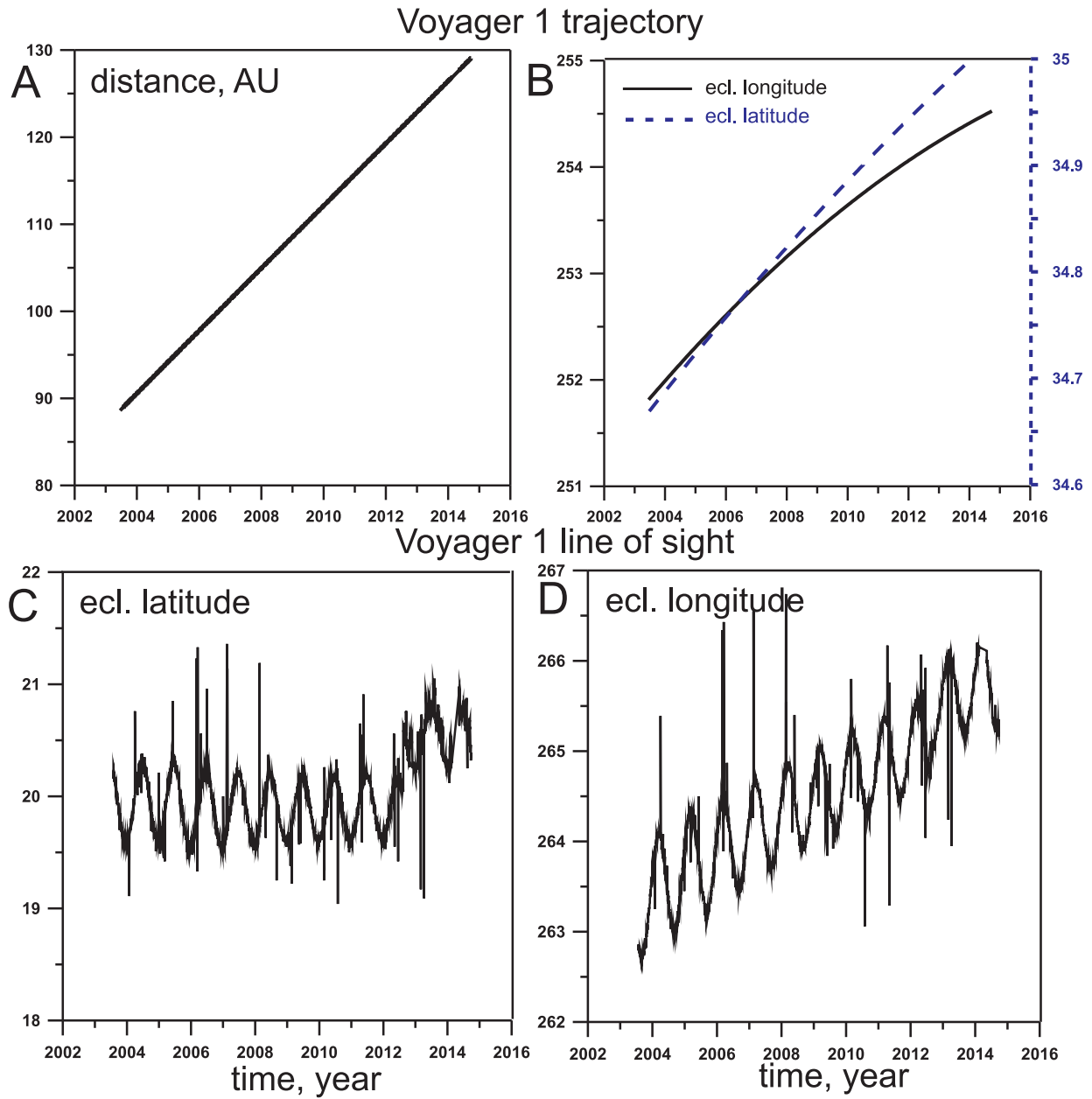}
\caption{Trajectory of Voyager~1 in 2003-2014: A. heliocentric distance; B. ecliptic longitude (solid curve) and latitude (dashed curve).
Voyager~1/UVS look directions after 2003: ecliptic latitude (C) and longitude (D). }
\label{traject}
\end{figure}

During the 11 years from 2003 to 2014 Voyager~1 has moved from $\sim$90~AU to 130~AU (see Fig.~\ref{traject}~A), its ecliptic
longitude has changed from 252$^{\circ}$ to 254.5$^{\circ}$ and ecliptic latitude has changed very slightly from 34.7$^{\circ}$ to 35$^{\circ}$ (see Fig.~\ref{traject}~B). This direction of motion is inclined at 30$^{\circ}$ to the upwind direction (see Fig.~\ref{interface}).
The UVS instrument has two ports, the main (airglow) and occultation ports. The total measured intensity is a sum of signal from both ports.
Contribution of the occultation port to the total intensity is about 8~\%.

Directions (in ecliptic J2000) of the UVS main port's LOSes are presented in Fig.~\ref{traject}~C and D. Here we exclude some irregular directions and consider
 LOSes with latitudes 19$^{\circ}-21.5^{\circ}$ and longitudes 262$^{\circ}-267^{\circ}$.
Periodic annual variations of the LOS direction are caused by changes of the spacecraft's orientation according to Earth's motion around the Sun.
The occultation port's view direction is always inclined to the main port at about 25$^{\circ}$. Hereafter, in mentioning Voyager's LOSes we
 refer to the main port.

The upwind direction determined from the interstellar helium observations by the Ulysses and IBEX spacecrafts is at $\sim$255.4$^{\circ}$ ecliptic longitude
and $\sim$5-6$^{\circ}$ ecliptic latitude \citep{witte_2004, mccomas_etal_2015}. Therefore, the LOS of Voyager~1/UVS since 2003 is inclined from the upwind
direction at about 17$^{\circ}$.


\subsection{Data processing}

The data used in the present work were processed following the scheme developed by \citet{holberg_watkins_1989} for the Voyager 1/2 UVS data.
Voyager-UVS is a grating spectrograph which means that photons are collected in 128 different channels according to their wavelength.
The processing scheme removes dark counts generated by particles (dark plate correction), corrects for scattering effects inside the instrument and applies a flatfield correction
on the different channels (Fixed Pattern Noise -- FPN). The Lyman~$\alpha$ counts are obtained by summing the channels 70 to 78 of Voyager 1 UVS.
This scheme has also been used by \citet{hall_thesis_1992} and \citet{hall_etal_1993}. However, in the case of Voyager~1 UVS, it must be noted that the flatfield correction
that we are using is different from the one used by \citet{hall_thesis_1992}. As noted in the annex of \citet{hall_thesis_1992}, he used a set of values denoted FPN 49 for Voyager 1, derived after
the Jupiter encounter. For our analysis of the UVS data, and all our works on the subject of interplanetary background data, we have used a more recent set of flatfield correction
denoted FPN 51 \citep{quemerais_etal_1995, quemerais_etal_2003, quemerais_etal_2013}.

In the case of the Lyman $\alpha$ line, it is straightforward to find a scaling factor between the count rates derived when using either FPN 49 or FPN 51. We have processed the whole dataset
used in this work using both flatfield correction coefficients. The ratio is almost constant, as can be expected since the line shape is very stable.
If the count rate found using FPN 49 and FPN 51 are denoted $count(FPN49)$ and $count(FPN51)$, correspondingly, we find that
\begin{equation}
count(FPN51) = 1.39 \times count(FPN49)
\end{equation}


Uncertainties for counts rate ($\sigma_{CR}$)
are computed based on the following formula, where $\Delta t$ is the integration time, N is the total number of counts and $CR=N/\Delta t$:
\begin{equation}
 \sigma_{CR} =\frac{\sqrt{ 2 \cdot N}}{\Delta t},
\end{equation}
the factor of 2 in the numerator is intended to account for a single photoevent sometimes producing counts in two or more adjacent channels,
leading to a situation in which not all counts are independent of one another.

It is very important to remove from the Lyman~$\alpha$ signal any contamination due to impacts of cosmic rays (solar or Galactic) or particles from the spacecraft radio-isotope thermoelectric generator (RTG) as they appear as an additional background spread over the whole bandwidth. The particle background spectrum was measured during dedicated observations
and was found to be very stable over the years, with very little channel to channel variation \citep[see supporting online material in][]{lallement_etal_2011}.
\citet{lallement_etal_2014} have studied in detail the temporal evolution of the Voyager~1 UVS background since 1993.
They found that there is a perfect correlation between the UVS background and the Cosmic Ray Subsystem (CRS) signal from energetic particles with energy above 70 MeV (see Fig.1-2 in \citet{lallement_etal_2014} and Fig.~\ref{data}~A for an illustration). They concluded that the UVS background is mostly due to impact of GCRs. To subtract this background from the data we follow the procedure developed by \citet{lallement_etal_2011, lallement_etal_2014}.
Namely, for each measurement, we determine the counts
obtained in two spectral regions without interplanetary emission, one shortward (channels 55-65) of 1216~${\rm \AA}$ and one longward (channels 85-95). We then scale the particle background reference spectrum to these two
spectral region. The resulting background spectrum is subtracted from the observed data.
This procedure ensures that there is no remaining contamination by dark counts, cosmic rays or particles from the RTG.
The final mean spectra (averaged over 2004-2012 and 2012-2014) are shown in Figure~\ref{spectrum} after we have applied descattering and flatfielding
as defined by \citet{holberg_watkins_1989}.

Fig.~\ref{spectrum} also shows that the spectrum is not affected by light from EUV stars in the vicinity of the line of sight. Such emissions have been observed some times as pointed out
by \cite{quemerais_etal_1995} (see figure 6 and figure 7 in that paper), but it was not the case for our data after 2003. As a matter of fact, the line of sight was chosen to avoid such
contamination.

A broadening of the bottom part of the spectrum (the wings of the line) is caused by the shape of the line spread function of the instrument. UVS is a low resolution spectrograph and consequently the line spread function (that is the spectral response to a monochromatic line) is strongly broadened. Also, the response is not a Gaussian, but instead the core part of the spectrum is widened at the bottom (due to instrumental effects, e.g., non specular diffusion on the optical elements plus some smearing after the micro-channel plate on the detectors). However, the shape of the line spread function is constant (ratio of the count rate over 3 central channels divided by the sum over 11 channels is very stable with time), therefore it does not influence the results.


Finally, it should be noted that during the period of interest for our analysis, Voyager 1 UVS suffered the following changes in temperature due to adjustments of power usage aboard Voyager 1:
\begin{itemize}
\item on 2005/01/26, The Voyager~1 UVS temperature dropped from -27 C to -55 C, accompanied by a change in the width of the H Lyman~$\alpha$ line. No change in count rate was observed.
\item on 2011/12/05, The IRIS replacement heater was turned off. The UVS temperature dropped by about 27 C. Within 10 days of this change, it was observed that the position and amplitude of
the H Lyman~$\alpha$ line had changed. The line center shifted by one pixel in the longward direction and the amplitude dropped by 10\% . Additionally,
 some pixels showed a strong increase in noise level but this did not affect the pixels used for the H~Lyman~$\alpha$ line measurements.
\end{itemize}

We have corrected the data obtained after the end of 2011, to match both count rate and line position as they were before the change in temperature. To do that, we averaged one month
of data before and after the change, and derived the position shift and scaling factor.

The average spectral shape obtained after 2012 is shown as a thin line in Figure \ref{spectrum}. The shift of the Lyman~$\alpha$ line by one channel is visible. We also notice that
the channel to channel response variation, which is very stable before the heater was turned off, changed after the temperature changed. For instance a peak appeared around 1000~${\rm \AA}$.
This is caused by a change in the response of the noise discriminators. But this change also appears to be very stable and does not affect the Lyman~$\alpha$ channels.

In this work, we investigate variations of Lyman~$\alpha$ intensity as a function of distance between the observer and the Sun.
Voyager~1/UVS data (original and averaged over 27 days) are shown in Fig.~\ref{data}~B.

\subsection{Correction for the solar Lyman~$\alpha$ flux}

The intensity of the backscattered solar Lyman~$\alpha$ emission in the heliosphere and beyond depends on the exciting solar Lyman~$\alpha$ flux.
The solar flux varies significantly at different time scales. Therefore, in order to use the UVS data to remotely sense temporal and spatial variations of the ISH distribution,
correcting the measured Lyman~$\alpha$ intensities for the solar flux is necessary.

The solar Lyman~$\alpha$ flux has been measured at the Earth orbit for decades \citep[e.g.][]{woods_etal_2000}. These measurements show
long-term solar cycle variations due to changes of solar activity, and short-term 27-day modulations due to rotation
of the Sun as it is seen from the Earth orbit.

We use the Composite Lyman~$\alpha$ time-series collected by the LASP Interactive Solar
IRradiance Data center (\url{http://lasp.colorado.edu/lisird/lya/}) providing the integrated solar Lyman~$\alpha$ flux with a time resolution of one day.
We note that
ISH atoms scatter solar photons only from the core of the Lyman-$\alpha$ line (due to a small Doppler shift from the line center corresponding to a typical
atom's velocity of $\sim$30~km/s). To transform the integrated Lyman~$\alpha$ flux to the required flux at the line center we use an expression provided by
\citet{emerich_etal_2005} that is based on direct measurements of the solar Lyman~$\alpha$ profile with SOHO/SUMER \citep{lemaire_etal_2002, lemaire_etal_2005}.

The intensity of the backscattered solar Lyman~$\alpha$ radiation obtained close to the Sun (e.g., at the Earth orbit by SOHO/SWAN) also shows 27-day modulations, because
the intensity is proportional to the local solar flux coming from the hemisphere of the Sun facing the observer.
However, \citet{quemerais_etal_1996a} and \citet{pryor_etal_2008} have shown that these modulations are damped significantly for an observer located quite far from the
Sun. This is a result of multiple scattering. At large distances (more than $\sim$50 AU) from the Sun the major portion of heliospheric Lyman~$\alpha$ emission
is due to solar photons scattered more than once. Therefore, photons detected by Voyager can originate not only
from the hemisphere of the Sun facing the spacecraft but rather from the whole surface of the Sun. Therefore, to correct Voyager's intensities obtained at each date we
average the LASP solar Lyman~$\alpha$ flux over 27 days ($\pm$13.5 days from the date of observations).
The original daily Lyman~$\alpha$ flux at the line center and the averaged values are presented in Fig.~\ref{data}~C.

Voyager~1/UVS data divided by the averaged solar Lyman~$\alpha$ flux at Earth are shown in Fig.~\ref{data}~D. It is seen that there is a very surprising period (2003-2009, 90-110~AU) when
 the corrected intensity is almost constant and then it decreases monotonically until the end of 2014.
The monotonic decrease of the measured brightness in 2010-2014 could be caused by a gradual degradation of the spectrograph sensitivity. Since 2003, when the Voyager~1 scan platform movements were stopped, there has been no way to check the instrumental calibration factor independently. However, there are two factors that suggest that the instrument remained stable and that the decrease is not due to an instrumental change. First, over the whole period, the dark count levels on the detector closely match the values obtained by cosmic ray experiments \citep{lallement_etal_2014}. Second, the decrease seen in the data starts before the last temperature drop which is a good indication that the two events are not correlated. Note that, whatever caused the decrease that occurred after 2010, it does not affect our study of the interesting ``stability'' period from 2003 to 2009, which is the main subject of this work.

 We analyze the data obtained in 2003-2014 using
results of the state of the art numerical model of the ISH distribution and radiative transfer described briefly in the next section.

\section{Numerical model}
\label{model}

In this work we use the same numerical model that was previously applied for our analysis of Voyager~1/UVS data obtained in 1993-2003
\citep{katushkina_etal_2016}. The ISH distribution is taken from results of a global 3-D stationary kinetic-MHD model of the SW/LISM interaction
\citep{izmod_alexash_2015}. This model takes into
account both heliospheric and interstellar magnetic fields (IsMF), heliolatitudinal variations of the solar wind parameters, and includes a kinetic description for hydrogen atoms.
Most of the calculations in this work use the base heliospheric model with the following LISM parameters:
number density of protons is $n_{p,LISM}=0.04$~cm$^{-3}$, number density of H atoms is
$n_{H,LISM}=0.14$~cm$^{-3}$, velocity of the interstellar wind is $V_{LISM}$=26.4~km/s and its direction is taken from the Ulysses interstellar neutral He data analysis reported by \citet{witte_2004} i.e. ecliptic (J2000) longitude 75.4$^{\circ}$ and latitude -5.2$^{\circ}$,
LISM temperature is $T_{LISM}$=6530~K, the IsMF strength is $B_{LISM}$=4.4~$\mu G$, angle between $\textbf{B}_{LISM}$ and $\textbf{V}_{LISM}$ is 20$^{\circ}$,
$(\textbf{B},\textbf{V})_{LISM}$-plane coincides with the Hydrogen Deflection Plane (HDP)
proposed firstly by \citet{lallement_etal_2005} and then slightly corrected in \citet{lallement_etal_2010}. A choice of these LISM parameters
is justified by \citet{izmod_alexash_2015}. We also perform several calculations with different LISM parameters as specified in the text.

Basically, two populations of the interstellar hydrogen are considered: primary atoms, which directly penetrate to the heliosphere from the LISM, and
secondary atoms, which are created by charge exchange between the primary atoms and interstellar protons outside the heliopause.
The most interesting feature of the ISH distribution at the heliospheric boundary is the hydrogen wall -- an increase of the ISH number density
outside the heliopause due to accumulation of slow secondary interstellar atoms created by charge exchange of the primary atoms with decelerated interstellar protons
\citep{baranov_etal_1991, bm1993}. Lyman~$\alpha$ photons scattered by slow atoms at the hydrogen wall have a Doppler shift
relative to the typical interstellar atoms in the heliosphere. Therefore, emission backscattered at the hydrogen wall is less scattered
inside the heliosphere and should be visible
from a much larger distance from the scattering point than the normal heliospheric glow.
\citet{katushkina_etal_2016} have shown that emission from the hydrogen wall can be identified in the Voyager~1/UVS data
obtained in the scans-regime in 1993-2003. In principle, the
H wall should lead to a flattening of a radial dependence of Lyman~$\alpha$ intensities measured in the distant heliosphere.

For one specific test calculation we took into account the heliospheric population of energetic hydrogen atoms (usually, this population is
called population~2, see, e.g. \citet{quemerais_izmod_2002}).
These atoms are created in the inner heliosheath region (between the HP and TS) by charge exchange of interstellar H atoms and solar wind protons.
Heliospheric atoms are much hotter than interstellar ones.
Due to their high temperature ($\sim10^6$~K) these atoms also provide backscattered photons with significant Doppler shift.
The number density of the heliospheric atoms in the base model is two orders of magnitude smaller than that of the interstellar atoms.
\citet{quemerais_izmod_2002} have shown that this hot component of the heliospheric hydrogen leads to significant broadening of the
backscattered Lyman~$\alpha$ line and provides about 5~\% of the total intensity at 1~AU from the Sun.


 The considered model of the ISH distribution is stationary, i.e. we ignore possible effects of the solar cycle on the hydrogen distribution.
 This is a plausible assumption because, as shown by, e.g., \citet{izmod_etal_2005} temporal variations of the hydrogen parameters far from the Sun
 are less than 10~\%.

To calculate the Lyman~$\alpha$ intensity seen by Voyager~1/UVS we used the radiative transfer code developed by \citet{quemerais_2000}.
The code accounts for multiply scattered photons, which provide the major contribution to the total intensity at large heliocentric distances.
We took into account actual positions and LOSes (for both main and occultation ports) of Voyager~1.
The general assumptions of the radiative transfer model are the following (the same as were used by Qu\'{e}merais \& Izmodenov, 2002):
\begin{itemize}
  \item The solar photons are randomly generated following the solar H Lyman~$\alpha$ line profile of Lemaire et al. (2005). The
source is spatially isotropic.
  \item The local distribution of hydrogen atoms is represented by
the sum of two components: primary and secondary. At each point inside the computational region the local distribution of
each component is described by 7 values: a number density,
a mean velocity vector and three temperatures which are the temperatures of the velocity distribution projected on the three
directions of the local frame. These parameters are calculated in the frame of the global heliospheric model.
  \item The scattering process is computed following Angle
Dependent Partial Frequency Redistribution \citep{quemerais_2000}.
  \item The scattering redistribution function includes the phase
function given by Brandt \& Chamberlain (1959).
\end{itemize}

In this work we use the most realistic ``Angle Dependent Partial Frequency Redistribution (ADPFR)'' frame for modeling of scattering process instead of simplified Complete Frequency Redistribution (CFR). A detailed description of the algorithm can be found in \citet{quemerais_2000}.  We also adopt realistic parameters of the hydrogen distribution including local velocity vectors and three kinetic temperatures taken from the results of the global heliospheric model. The only important assumption is that the local hydrogen velocity distribution function of both primary and secondary H atoms is assumed to be Maxwellian with three different temperatures at each point. In reality, the distribution function of the secondary hydrogen atoms might be asymmetric due to charge exchange process at the heliospheric boundary and inside the heliosphere \citep[see, e.g.,][]{izmod_etal_2001}. However, the shape of the hydrogen velocity distribution function mainly affects the spectral properties of the backscattered Lyman~$\alpha$ profile but not the considered intensities. Therefore we do not expect any significant uncertainties related to this assumption. The same model was previously used for modeling of the Voyager/UVS data obtained in 1979-1995 and the model results show a good agreement with the Voyager data \citep{quemerais_etal_2013}. The Monte Carlo computation involved 10$^9$ photons, and corresponding statistical uncertainties are not more than few percent \citep{quemerais_izmod_2002}.

\section{Comparison of the model results with observations}
\label{comparison}

 In this section we compare the Voyager~1/UVS data with our simulations performed in the framework of the model described above.
Fig.~\ref{comp_models}~A shows the normalized intensity divided by the solar flux as a function of heliocentric distance.
The data are shown in units of counts per second and not Rayleighs. The calibration of Voyager-1/UVS at Lyman~$\alpha$ is under revision
as shown by recent studies \citep{quemerais_etal_2013, ben-jaffel_etal_2016}.
The calibration factor derived by Hall (1992) corresponds to data that were processed with FPN 49. The FPN correction is essentially a flatfield correction
for the various channels of the UVS detector (See Holberg and Watkins, 1989; Hall, 1992). The data that we are using in this work were processed
using FPN 51. This leads to a change of the count rate at Lyman~$\alpha$ by a factor of 1.4 (as mentioned in Section~2.2). Other wavelengths are not affected in the same way.
Therefore, the calibration factor given by Hall (1992) should be corrected by the same factor. This leads to a value of 218/1.4 Rayleigh per counts per second,
when using FPN 51 while processing data.
However, as shown by \citet{quemerais_etal_2013}, this calibration factor does not agree with our model of the interplanetary background.

We are currently working on a multi-spacecraft comparison of data that will show that the calibration factor of Voyager~1/UVS and Voyager~2/UVS
at Lyman~$\alpha$ should be revised. The results of the calibration analysis will be published in separate paper.
In this work we are interested only in the radial dependence of intensity and its gradient and do not consider absolute values of intensities.



 It is seen that the model intensity decreases monotonically, while the data show almost constant intensity at 90-110~AU.
 In addition to results of the base model, Fig.~\ref{comp_models}~A presents the results for two additional models, which differ from the base
 model in the LISM hydrogen and proton number densities, see Table~\ref{table_1}.
 The height, width and location of the hydrogen wall are substantially different for these three models.
The same models were considered before by \citet{katushkina_etal_2016}, and it was shown that they give significantly different results for
the upwind to downwind ratio of Lyman~$\alpha$ intensities, which were also measured by Voyager-1 in 1993-2003. However, Fig.~\ref{comp_models}~A
shows that all three models give approximately the same results
for the radial dependence of the intensity measured by Voyager in 2003-2014, and this dependence is qualitatively different from the Voyager data.
Namely, all models predict a monotonic decrease of intensity with distance from the Sun.

Actually, this behavior of the model intensity is not surprising. It is linked to the fact that for these observations, the line of sight direction
is almost co-linear with the position vector of the observer. Appendix~A presents a theoretical consideration of the dependence of intensity on distance
from the Sun when the LOS direction and the position vector are co-linear.
It is shown that in the simplified case without absorption of backscattered photons, the Lyman~$\alpha$ intensity measured by Voyager~1 should certainly decrease with distance from the Sun,
whatever the hydrogen density profile.

\citet{hall_etal_1993} have studied the dependence of intensities on the observer's
distance from the Sun. They showed that in the frame of the classical hot model \citep[see, e.g.,][]{lallement_etal_1985} with a constant hydrogen
 number density in the outer heliosphere the Lyman~$\alpha$ intensity in the upwind direction is proportional to $1/r$ in the optically thin case (when only once scattered photons are
considered) and tends to $1/r^2$ in the limit
of the optically thick case (when multiply scattered photons dominate). They also have noticed that a positive gradient of hydrogen number density
should lead to smaller rates of intensity decrease in both optically thin and thick limits. \citet{quemerais_2000} have calculated
intensities in the upwind direction for an observer located at 80-100~AU from the Sun and determined the radial coefficient $\alpha$
such as $I(r)\sim r^{\alpha}$. They found that for the classical hot model $\alpha=-0.99$ in the optically thin case and $\alpha=-1.46$ in the
radiative transfer case (when multiple scattering is taken into account). \citet{quemerais_etal_2003} did the same for the more realistic model of
the hydrogen distribution including the hydrogen wall. They found $\alpha=-0.79$ in the optically thin case and $\alpha=-1.06$ in the
radiative transfer case. These results agree with predictions of \citet{hall_etal_1993}.
\citet{fayock_etal_2015} have presented their model
intensities (as functions of distance) for the look directions of Pioneer~10 and Voyager~2 spacecraft. Their intensities also decrease monotonically for
distances from $\sim$7 to 150~AU.

Thus, all available models of the H distribution in the heliosphere imply a monotonic decrease of intensity with heliocentric distance,
while the Voyager~1/UVS data in 2003-2009 show qualitatively different behavior.
Two possible explanations can be proposed. The first one is that we need to change some model parameters (e.g. solar wind and LISM boundary conditions)
to obtain an appropriate hydrogen distribution. The second is that we need to incorporate some new physics in the model.

To check the first possibility we perform four test calculations with artificially changed hydrogen distributions in the heliosphere obtained from
the base model, as follows. Models A and B relate to the changes of configuration of the hydrogen wall.
In Model~A the hydrogen number density
in the hydrogen wall is increased by a factor of 5 compared with the base model. In this case the gradient of the number density in the hydrogen wall is larger and in accordance with
\citet{hall_etal_1993} a rate of decrease of $I(r)$ should be smaller. In Model~B the hydrogen wall is closer to the Sun (all distances
in the model are scaled by a factor of 0.5). In principle, it also should increase the role of the H wall at the distances of Voyager-1 observations,
and therefore, change the gradient of $I(r)$. Models~C and D deal with different populations of H atoms. In Model~C we artificially increase the
separation between energy distributions of the primary and secondary interstellar hydrogen atoms. Namely, in this model
all velocities of primary atoms are multiplied by a factor of 2, all velocities of secondary atoms are multiplied by a factor of 0.1 and all temperatures
of secondary atom are multiplied by 0.5. As a result, the velocity distribution functions of the primary and secondary atoms become non-overlapping (see Fig.~\ref{modelC}).
In this case Lyman~$\alpha$ photons scattered by the secondary atoms at the hydrogen wall should be less absorbed
in the heliosphere and so more visible. Model~D includes a heliospheric population of H atoms with number density increased by a factor of 10
compared with the base model.

Fig.~\ref{comp_models}~B presents the results of these four test models together with the base model results and the Voyager data.
It is seen that although in the test models we change the hydrogen distribution quite significantly, the effect on the radial dependence is
quite small. All models still predict a monotonic decrease of intensity. Therefore, we can conclude that any intuitive changes of hydrogen distribution
do not help to explain the Voyager data.

In the next section we consider two tentative scenarios, which could explain the Voyager~1/UVS data at 90-110~AU from the Sun.

\section{Two scenarios}
\label{scenario}

\subsection{Additional dense layer of hydrogen atoms near the heliopause}

The first idea is that an additional dense layer of H atoms may exist in the inner or outer heliosheath close to the heliopause.
Lyman~$\alpha$ photons scattered by H atoms in the layer should have significant Doppler shift to be not absorbed inside the heliosphere.
Therefore, for an approaching observer the layer can be considered just as a source of constant backscattered emission, and
 when the observer reaches the layer the intensity starts to decrease due to strong absorption inside the layer (see illustration in Fig.~\ref{layer_scheme}~A).

Several physical reasons for the layer formation can be considered.
For example, \citet{lallement_etal_2014} proposed
the physical mechanism of layer formation in the inner heliosheath by enhanced charge exchange between suprathermal pick up ions (PUIs) and interstellar neutrals. \citet{merav_etal_2012} have considered a Flow Transition Region upstream the heliopause, where the flow abruptly turns and the radial velocity becomes near zero. Charge exchange in this region could also produce an enhanced number density of hydrogen with Doppler shifted velocities.
In principle, the hydrogen wall outside of the heliopause could contain a specific population of hydrogen atoms with positive velocities created by
charge exchange with energetic particles.

In this work we do not specify the origin of the layer. We included the layer artificially and find its parameters that produce appropriate
additional component of the Lyman~$\alpha$ emission scattered by such a layer. More investigations are needed in order to define the layer
self-consistently in the frame of the global heliospheric model.

The H number density profile assumed in the layer is shown in Fig.~\ref{layer_scheme}~B. This is just an artificial model with constant hydrogen parameters
inside the layer. We found that appropriate results can be obtained with the following parameters of the layer:
location at about 120-160~AU from the Sun, H number density is very high $\sim10\,cm^{-3}$, z-component of velocity is $V_{z,H}\sim+50$~km/s,
temperature is about 10000~K. Such a velocity provides enough Doppler shift. Fig.~\ref{layer_res} presents the results: the main component of intensity
 arises from the original hydrogen distribution discussed above,
an additional component from the layer, and the total intensity that is their sum. We see that the additional intensity from the layer is
several times larger than the nominal one for an observer located at $r\geq90$~AU. Therefore, the total intensity is almost constant
at 90~AU$\leq r\leq$115~AU and then sharply decreases. Comparing with the Voyager data shows that
 the agreement is strongly improved from 90 to 120-125 AU, but beyond 125 AU this model predicts a rapid decrease of the signal, in strong contradiction with the data.

The emission from such a dense layer would be observable from the inner heliosphere. Indeed, the brightness of such a layer is of the order
of 30 to 60 Rayleighs and is strongly Doppler-shifted away from the interplanetary line center. As its velocity is positive, i.e. flowing away
from the Sun, the corresponding spectral feature would appear on the red wing of the interplanetary line. This means that for an instrument
in low Earth orbit, the spectral feature would be superimposed on the strong geocoronal line \citep{clarke_etal_1998} and thus impossible
to observe. Only a high resolution spectrograph away from the strong coronal emission of a planet could observe it.
Note also, that the hypothetical spectral feature reported by \citet{benjaffel_2000} is not appropriate here since it was on the blue wing
of the interplanetary line.

\subsection{Additional component of external non-heliospheric emission}

The second idea is to include a constant additive non-heliospheric component of Lyman~$\alpha$ emission. In principle it could be connected with the galactic
or ``extragalactic'' Lyman~$\alpha$ background or some other external source of isotropic radiation. The look direction of Voyager~1 after 2003 is close to the galactic plane although
not directly in it.
Previously several authors have suggested that the addition of a constant non-heliospheric background is needed for better fitting of Voyager
Lyman~$\alpha$ data, see \citet{hall_etal_1993}, \citet{quemerais_etal_1996b} and \citet{lallement_etal_2011}.

 Our test calculations with such an artificially added constant background showed that 25~R of additional emission
 provides the best agreement with the Voyager data.
Such an additional brightness is far above the Galactic contribution expected to follow the relationship between H~$\alpha$ and Lyman~$\alpha$ established by Lallement et al. (2011) based on the UVS scans, which is only a few Rayleigh.
Fig.~\ref{add25R} shows the results of the comparison. As previously, this plot presents intensities normalized to the solar Lyman~$\alpha$ flux
at Earth orbit.
The non-heliospheric emission does not depend on the solar Lyman~$\alpha$ flux, and therefore we divide it by the solar flux in order to compare with the
normalized heliospheric intensities. The normalized additional emission increases with distance from the Sun at $\sim$80-100~AU
because of the decrease in solar Lyman~$\alpha$ flux. As a result, the total normalized intensity is constant at 90-110~AU in agreement with the Voyager~1/UVS data.
Therefore, this scenario is also a possible qualitative explanation of Voyager UVS data. Note that from the present analysis of Voyager~1/UVS data
in one direction we can not determine whether the additional emission is isotropic or concentrated in a specific region. In principle, both variants are possible. However, New Horizons/Alice cruise observations (see next section) suggest that this additional emission is restricted to the upwind hemisphere.

\section{New Horizons Cruise Observations}
\label{alice}

During its cruise to Pluto, the Alice ultraviolet spectrograph \citep{stern_etal_2008} has performed observations of the interplanetary medium. These observations have been
described by \citet{gladstone_etal_2013}. They consist of rolls around a fixed vector with the LOS longitude ecliptic covering 360 degrees while the latitude varies
from -30 to +30 degrees. The first roll was not complete and did not cover the upwind hemisphere. The second and fourth rolls were complete and were obtained at
11.3 AU and 17.0 AU from the Sun, respectively. We have run our model for the Alice observations in exactly the same way as the one described in section~\ref{model}.

The results for the roll ACO-2 at 11.3 AU are shown in Fig.~\ref{aco2} and the ones for the roll ACO-4 at 17.0 AU are shown in Fig.~\ref{aco4}. The data are shown in Rayleigh.
The solar flux used in the computation is the one obtained following the scheme described in section \ref{data}. The model is in excellent agreement with the data except close to the
upwind direction. The excess of brightness close to the upwind is of the order of 20 Rayleighs and is very similar for both rolls ACO-2 and ACO-4 which suggests that this excess is not
varying as the spacecraft moves outward.

These observations are an independent confirmation that there is an excess of Lyman $\alpha$ emission in the upwind direction and that this excess does not vary with distance.
As the ACO2 and ACO4 rolls did not go through the same point as the Voyager 1/UVS data the actual value in Rayleigh for the excess could be slightly different, depending on how close
the LOS is to the upwind direction. Also, the Alice data could be compatible with both hypotheses described above.

\section{Discussion and conclusions}
\label{discussion}

In this paper we present for the first time the Voyager~1/UVS data obtained in 2003-2014. These measurements were performed for a single line
of
sight toward the nose part of the heliosphere. The data (after normalization by the solar Lyman~$\alpha$ flux)
show a surprising flat period at 90-115~AU from the Sun, where the intensity is almost constant, that can not be reproduced and explained by existing global heliospheric models.
An additional source of emission is necessary to fit the data. We suggest two possible qualitative scenarios that can explain the data.

The first scenario is motivated by the idea of existence of a dense hydrogen layer near the heliopause.
One possible reason for forming such a layer was proposed by \citet{lallement_etal_2014}. They suggested that the charge exchange rate
is especially significant for the suprathermal PUIs in the inner heliosheath region due to their frequent gyrorotation,
small bulk velocity and large individual velocities. An enhanced charge exchange rate results in the creation of many new hydrogen atoms with
 individual velocities of the parent PUIs. These new atoms may accumulate and create a dense layer.
Our numerical simulations of the Lyman-alpha intensities coming from an artificially added layer show that quite large number densities ($\sim10\,cm^{-3}$) of H atoms in the layer with significant Doppler shifted velocities (+50 km/s) are needed to produce a more or less constant intensity in agreement with the Voyager~1/UVS data. Such hydrogen parameters are unusual indeed. We speculate about the potential creation of such a layer by extensive charge exchange near the heliospheric boundary. However, further investigations are needed for quantitative estimations of the effect.


The second scenario deals with extra-heliospheric emission, such as, e.g., a galactic or even extra-galactic Lyman~$\alpha$ background.
Galactic emission should be concentrated only in the galactic plane, while Voyager's look direction after 2003 is out of galactic plane at
about 17$^{\circ}$. We found that 25~R of external isotropic emission gives good agreement with the Voyager data.

Data obtained by Alice during the New Horizon cruise to Pluto \citep{gladstone_etal_2013} show that there is an excess of emission in the upwind hemisphere.
These data were obtained at 11.3 AU and 17 AU. The excess in both cases is close to 20 R. Such an excess over our model computations confirm
our findings based on the analysis of the Voyager 1/UVS data. Unfortunately, these observations do not bring new information on the origin of this excess.

Previously several authors speculated that additional emission is necessary to explain the Voyager data.
\citet{hall_etal_1993} stated that the quality of the fit to the Voyager data at 25-47~AU may be improved by including a constant additive intensity
of about 20-30~\% of that observed at 15~AU (i.e. about 70-80~R). \citet{quemerais_etal_1996b} showed that the Voyager data obtained during the scan-regime in 1993-1994 ($\sim$50-58~AU)
could be represented by the classical hot model of the hydrogen distribution only by adding 10-15~R of external (galactic) emission.
\citet{lallement_etal_2011} have shown that Voyager data show an excess of upwind emission relative to the results of the global heliospheric model.
They concluded that this excess is a trace of galactic brightness coming from the the Scorpius-Ophiuchus area. Model dependent analysis
allowed them to estimate this emission at 3-4~R.

Previously \citep{katushkina_etal_2016} we have shown that our state-of-art heliospheric model predicts a systematically larger ratio of Lyman~$\alpha$
intensities at the nose and tail directions compared to the Voyager~1/UVS data obtained in the scans regime in 1993-2003.
It was suggested that one possible solution of this problem is to increase the hydrogen wall by e.g. changing the number densities of the
LISM protons and hydrogen atoms. We also mentioned another possible solution, adding a constant emission for all lines of sight.
Now we perform a new fitting of the data in the frame of the base heliospheric model with an additional constant isotropic emission.
We find that 15-20~R of additional emission leads to a good agreement with the Voyager data in 1993-2003 (see Fig.~\ref{ratio}).
Thus, the additional constant emission allows us not only to explain the Voyager data after 2003, but also provides a better agreement with the model
results for the previous period of observations (1993-2003).

\begin{acknowledgments}

The data used here are from the
UVS data archive at the Lunar and
Planetary Laboratory, University of
Arizona. These data were derived
directly from Voyager telemetry
information contained in UVS
Experiment Data Records provided
by JPL to the UVS investigation
team. The data are available from
the authors upon request (Eric
Quemerais, eric.quemerais@latmos.
ipsl.fr). New Horizons/Alice data are available on-line - \url{https://sbn.pds.nasa.gov/holdings/nh-x-alice-2-plutocruise-v2.0/dataset.html}
(Stern, A., NEW HORIZONS Raw ALICE PLUTO CRUISE V2.0, NH-X-ALICE-2-PLUTOCRUISE-V2.0, NASA Planetary Data System, 2016).

We thank Dmitry Alexashov for providing us the hydrogen distribution obtained by him in the frame of the global heliospheric model.
We thank our anonymous referees for their useful questions, which help to improve the paper.

 O.K. acknowledges support
from CNES during her postdoctoral
position at LATMOS, France.

Global astrospheric/heliospheric modeling has been performed in the frame of Russian Science Foundation project 14-12-01096. Computation and analysis of the backscattered Lyman-alpha is supported by RFBR-CNRS project 16-52-16008.

Calculations of the ISH distribution were performed by using the Supercomputing Center of Lomonosov Moscow State University.

\end{acknowledgments}

\appendix

\section{Theoretical study of the radial dependence $I(r)$}

We consider the problem in more detail from a theoretical point of view. In general, the intensity measured at a fixed LOS by an observer
looking radially away and located at distance $r$ is characterized by the integral along the line-of-sight:
\begin{equation}
 I(r)=\int_{0}^{\infty} d\lambda \int_{r}^{\infty} \varepsilon_{\lambda}(s) exp(-\tau_{\lambda}(r,s)) ds,
\end{equation}
where $\lambda$ is the wavelength, $s$ is the coordinate along the LOS, $\varepsilon_{\lambda}(s)$ is the local emissivity
due to scattering at the point with coordinate $s$, and $\tau_{\lambda}(r,s)$ is the optical depth between the observer and scattering point.
To simplify notation let us consider the intensity for the chosen wavelength $I_{\lambda}(r)$ such as, $I(r)=\int_0^{\infty} I_{\lambda}(r)d\lambda$.
Then, let us consider two positions of the observer along the LOS: $r_1$ and $r_2$ (see Fig.~\ref{tau}). Therefore,
\begin{equation}
\begin{aligned}
 I_{\lambda}(r_1) &= \int_{r_1}^{\infty} \varepsilon_{\lambda}(s) exp(-\tau_{\lambda}(r_1,s)) ds, \\
 I_{\lambda}(r_2) &= \int_{r_2}^{\infty} \varepsilon_{\lambda}(s) exp(-\tau_{\lambda}(r_2,s)) ds.
  \end{aligned}
\end{equation}
Take into account that for any $s\geq r_2$:
\begin{equation}
\begin{aligned}
 \tau_{\lambda}(r_1,s) &= \tau_{\lambda}(r_1,r_2) + \tau_{\lambda}(r_2,s) \, \Rightarrow \\
 exp(-\tau_{\lambda}(r_1,s)) &= exp(-\tau_{\lambda}(r_1,r_2)) \cdot exp(-\tau_{\lambda}(r_2,s)).
  \end{aligned}
\end{equation}
Thus,
\begin{equation}
\begin{aligned}\label{intensity}
I_{\lambda}(r_1) &= \int_{r_1}^{r_2} \varepsilon_{\lambda}(s) exp(-\tau_{\lambda}(r_1,s)) ds + exp(\tau_{\lambda}(r_1,r_2)) \cdot \int_{r_2}^{\infty} \varepsilon_{\lambda}(s) exp(-\tau_{\lambda}(r_2,s)) ds  \\
&= \int_{r_1}^{r_2} \varepsilon_{\lambda}(s) exp(-\tau_{\lambda}(r_1,s)) ds + exp(-\tau_{\lambda}(r_1,r_2)) \cdot I_{\lambda}(r_2).
  \end{aligned}
\end{equation}

If $\tau_{\lambda}=0$ one has:
\begin{equation}
I_{\lambda}(r_1)=I_{\lambda}(r_2)+\int_{r_1}^{r_2} \varepsilon_{\lambda}(s) ds \geq I_{\lambda}(r_2).
\end{equation}
Hence, $I_{\lambda}(r)$ is a nonincreasing function for any kind of $\varepsilon_{\lambda}(s)$. And it could be constant at $r\in[r_1,r_2]$ only if $\varepsilon_{\lambda}(r)\equiv0$ at this interval.

\end{article}

\clearpage

\begin{table}
\caption{Model parameters and results \label{table_1}}
 \centering
\begin{tabular}{lcc}
\hline
No. & $n_{H,LISM}$\tablenotemark{a}, cm$^{-3}$ & $n_{p,LISM}$\tablenotemark{b}, cm$^{-3}$  \\
\hline
model~1 & 0.14 & 0.04  \\
model~2 & 0.18 & 0.06  \\
model~3 & 0.2  & 0.1   \\
\hline
\end{tabular}
\tablenotetext{a}{Number density of H atoms in the undisturbed LISM.}
\tablenotetext{b}{Number density of protons in the undisturbed LISM.}
\end{table}

\clearpage

\begin{figure}
\noindent\includegraphics[scale=0.6]{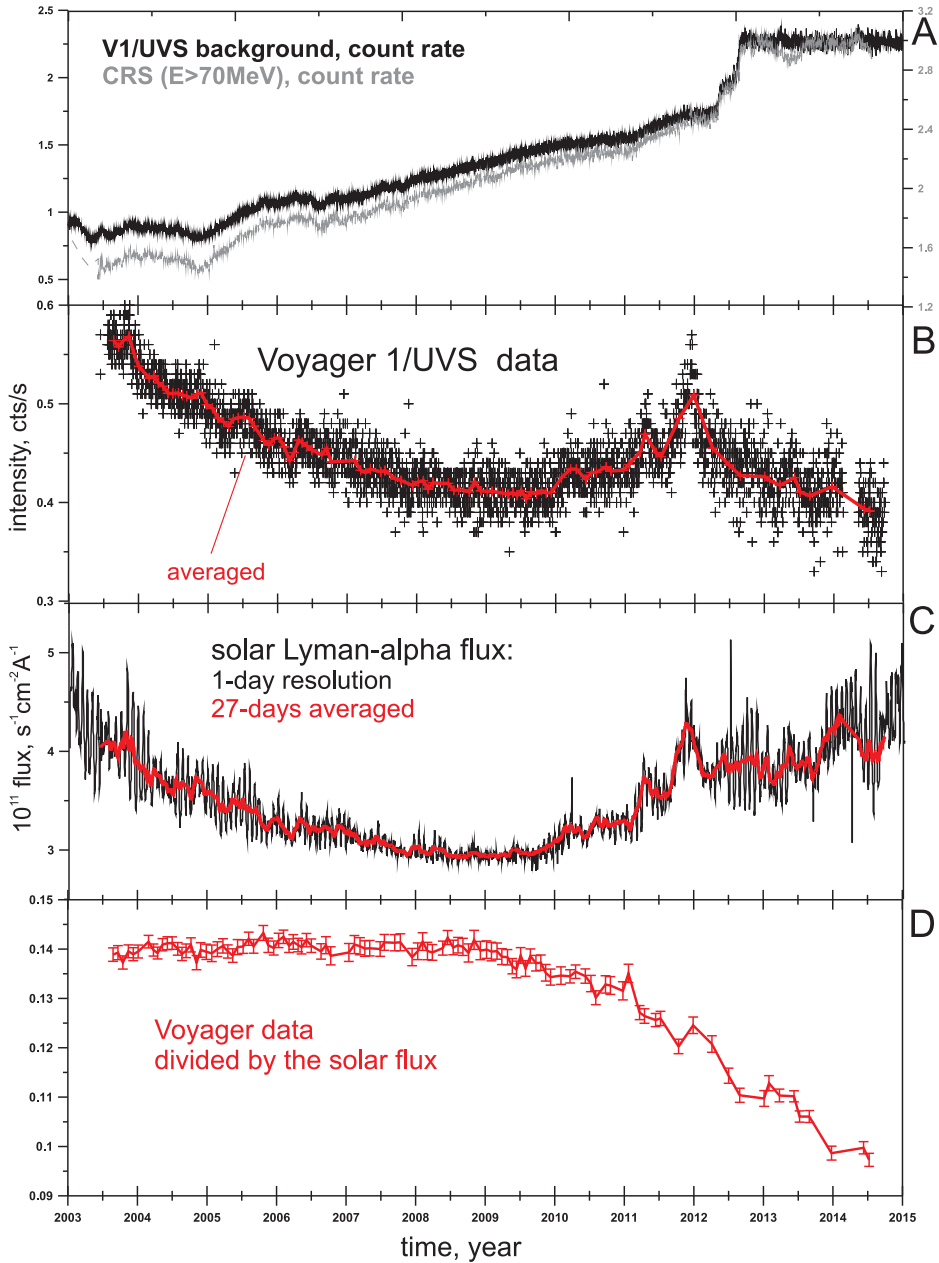}
\caption{A. Comparison of the UVS background and Cosmic Ray Subsystem (CRS) count rate for protons with energy above 70 MeV
(the one that matches very well the range of sensitivity of the UVS detector). B. Voyager~1/UVS Lyman~$\alpha$ signal measured in 2003-2014 obtained after
background subtraction (see Section 2.2 for details). Symbols correspond to original data and the red curve corresponds to
the data averaged over 27 days. C. Solar Lyman~$\alpha$ flux at the line center at Earth with one-day resolution and averaged over 27 days.
D. Voyager intensity divided by the averaged solar flux. }
\label{data}
\end{figure}

\begin{figure}
\noindent\includegraphics[scale=0.8]{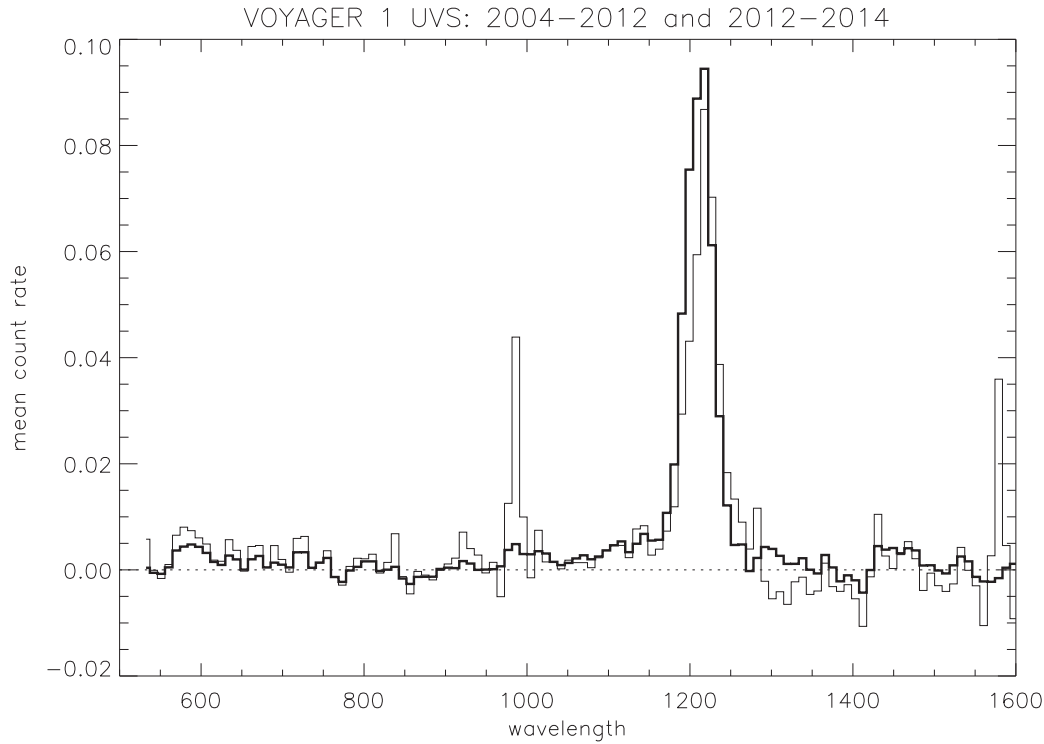}
\caption{Mean spectrum recorded by Voyager between 2004 and 2012 (thick solid line) obtained after background substraction (see Section 2.2 for details). The spectrum shape is very constant. The Lyman~$\alpha$
line is the only visible line (around 1216~${\rm \AA}$). This shows that there is no stellar contamination. The thin line corresponds to the spectrum shape
from 2012 to 2014. First we notice a shift of the Lyman~$\alpha$ line by one channel as pointed in the text. Second, there is also a change in the pixel
to pixel response which is due to a change in the noise discriminator of the channels due to the low temperature. These values are constant and do not affect
the count rate of the Lyman~$\alpha$ line (channels 70 to 80).}
\label{spectrum}
\end{figure}

\begin{figure}
\noindent\includegraphics[scale=0.8]{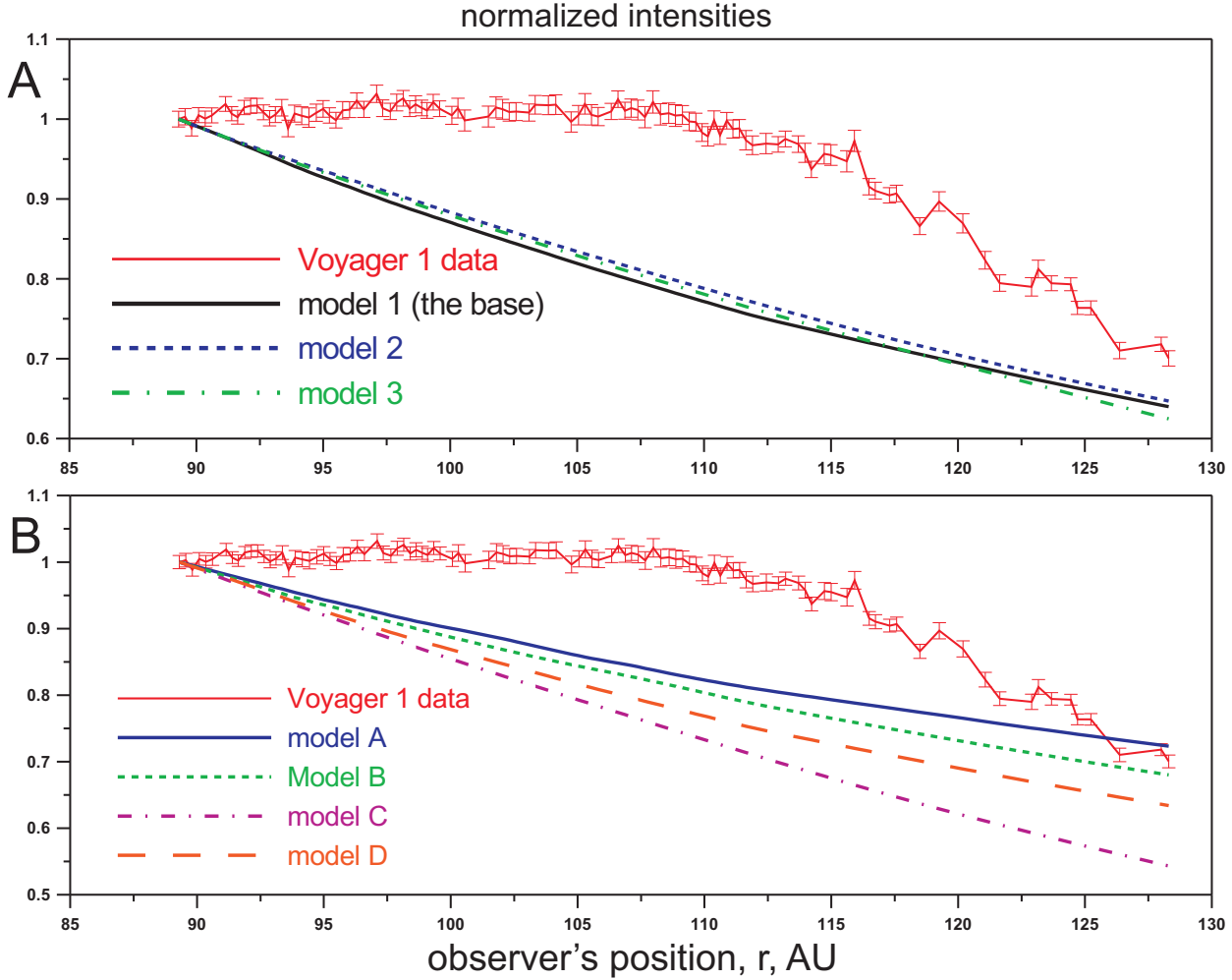}
\caption{Comparison of the model results with Voyager~1/UVS data. All intensities are divided by the solar flux and normalized to unity at
the first observational point (90~AU). A. Results of three models with different LISM parameters (see Table~1).
B. Results of four test model calculations (A-D)  are shown, see text for description of each test case.}
\label{comp_models}
\end{figure}

\begin{figure}
\noindent\includegraphics[scale=0.8]{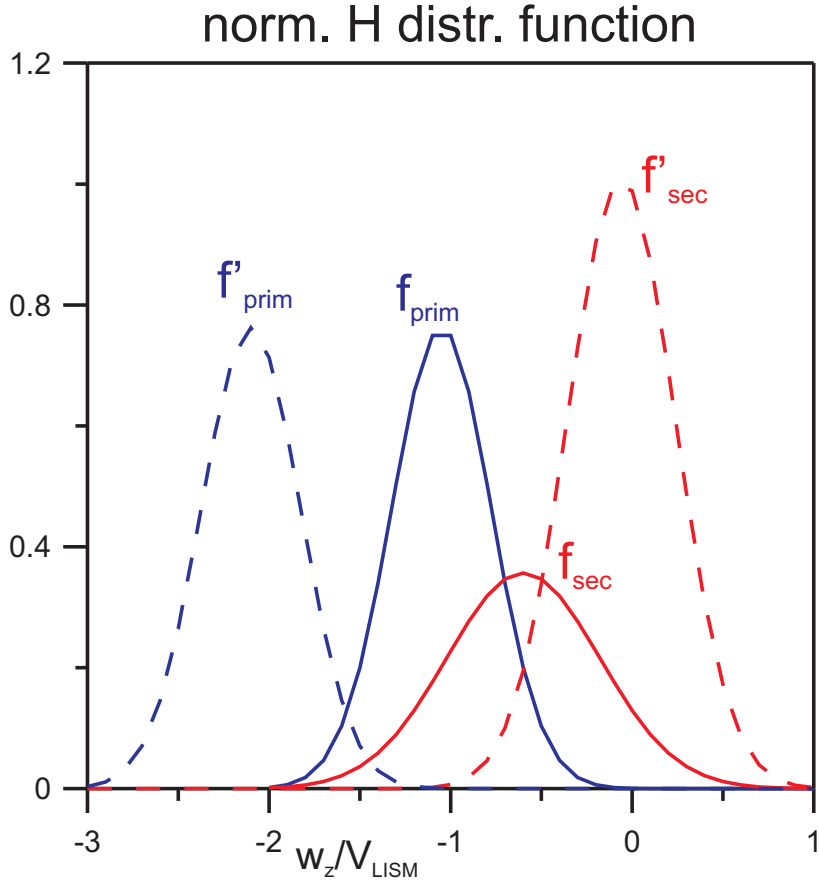}
\caption{Example of a hydrogen velocity distribution function inside the heliosphere. The index ``prim'' corresponds to the primary interstellar
atoms, the index ``sec'' corresponds to the secondary interstellar atoms. Solid curves show results of the base model,
where distribution functions of two populations overlap significantly. The dashed curves show results of the test Model~C, where
the bulk velocity $V_z$ of primary atoms is multiplied by a factor of 2, the bulk velocity of secondary atoms is multiplied by a factor of
0.5 and temperature of secondary atoms is multiplied by 0.5. In this test case the primary and secondary atoms are separated from each other.  }
\label{modelC}
\end{figure}

\begin{figure}
\noindent\includegraphics[scale=0.8]{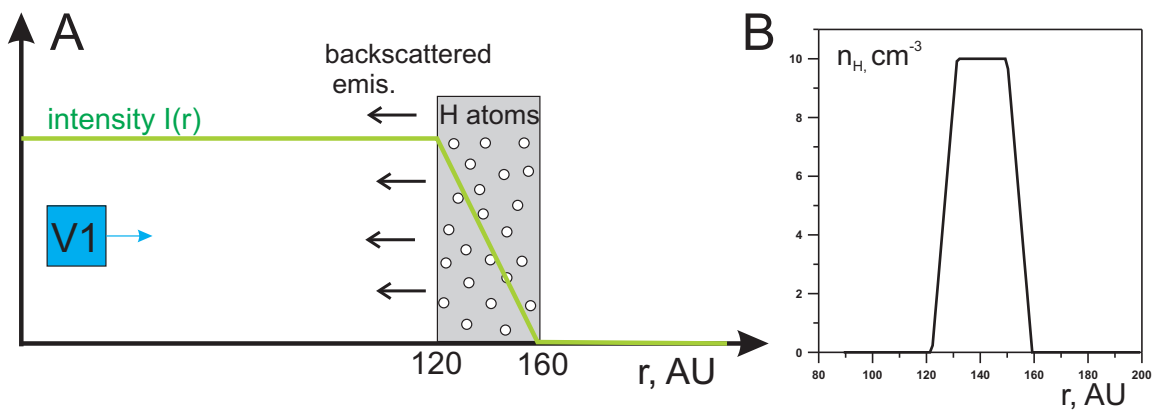}
\caption{A. Schematic illustration of Lyman~$\alpha$ intensity backscattered at a dense layer with significantly Doppler shifted H atoms.
The green solid curve shows the intensity from the layer that would be observed by spacecraft at different distances from the Sun.
B. Hydrogen number density profile at the layer.}
\label{layer_scheme}
\end{figure}

\begin{figure}
\noindent\includegraphics[scale=0.8]{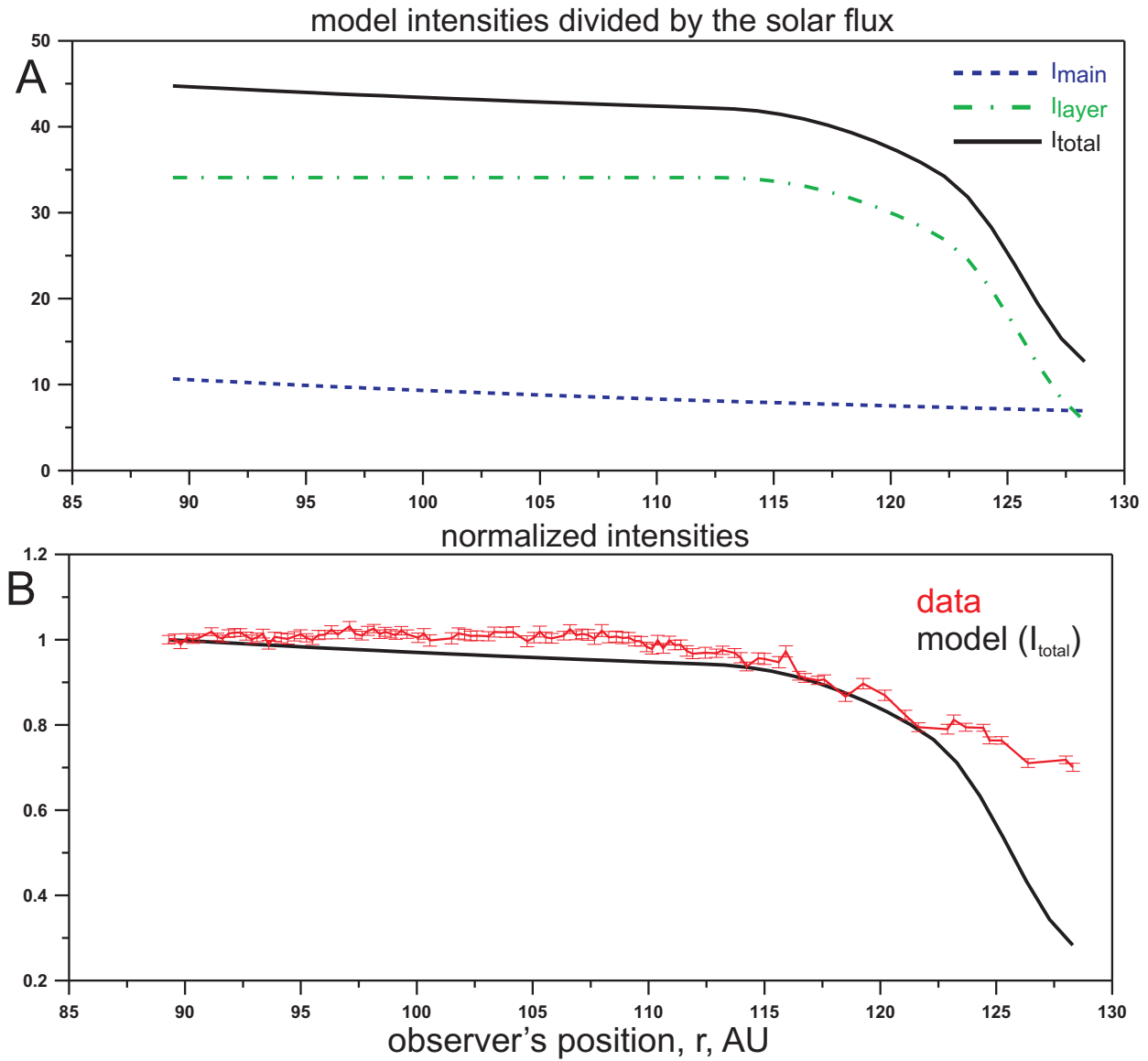}
\caption{First artificial scenario with an additional dense layer of atoms near the heliopause.
All intensities are divided by the solar flux.
A. Three components of the Lyman~$\alpha$ intensities are shown.
The dashed curve shows the main component of intensity obtained from the base model, the dashed-dot curve shows the additional intensity coming from a dense
layer of H atom near the HP, and the solid curve shows the total intensity, which is a sum of the two components.
B. Comparison of the normalized total intensity with the data. }
\label{layer_res}
\end{figure}

\begin{figure}
\noindent\includegraphics[scale=0.8]{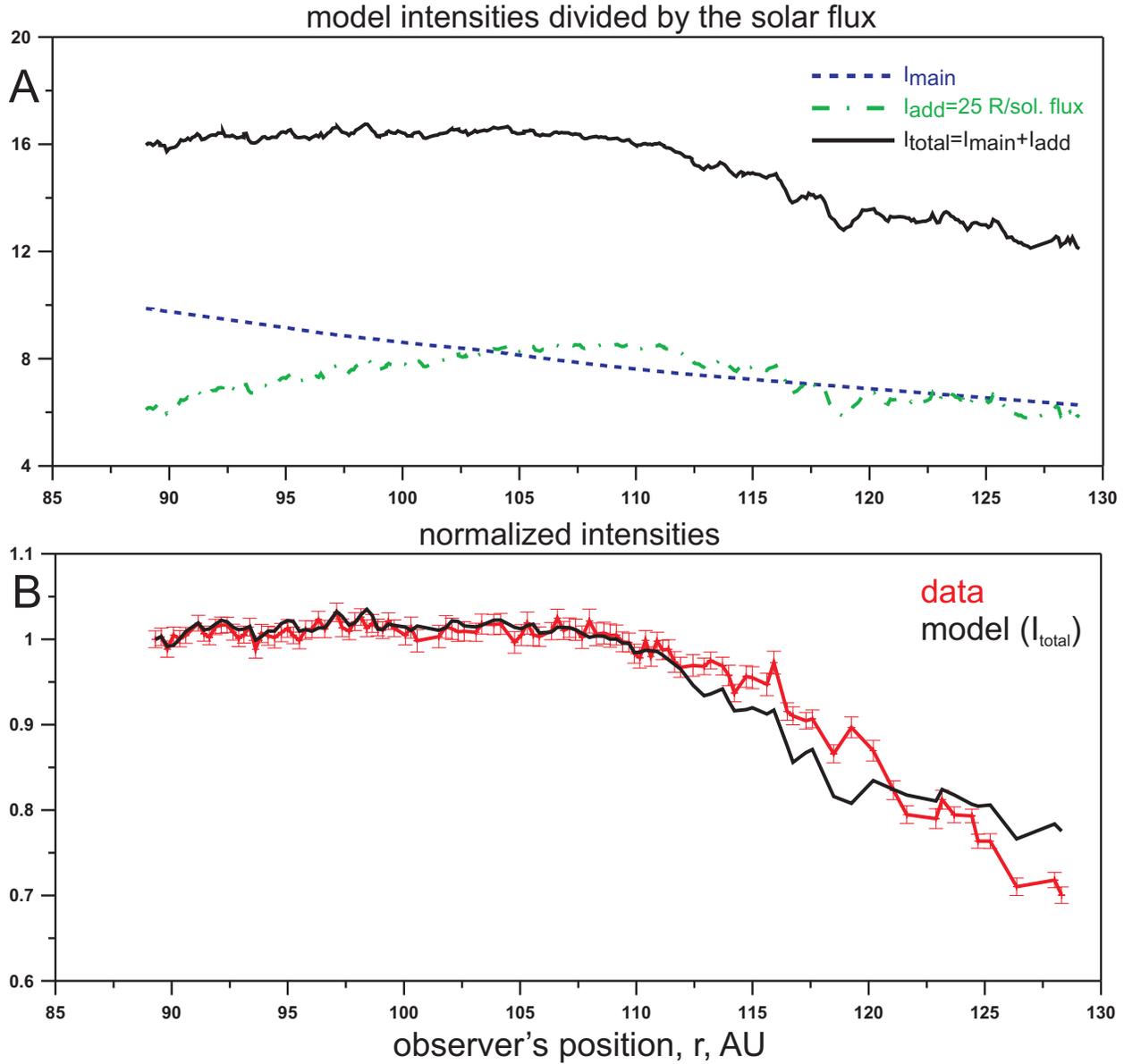}
\caption{Second artificial scenario with an additional 25~R of external (extra-heliospheric) uniform emission.
All intensities are divided by the solar flux.
A. Three components of the Lyman~$\alpha$ intensities.
The dashed curve shows the main component of the intensity obtained from the base model, the dashed-dot curve shows the additional intensity, that is a constant intensity
divided by the solar flux, and the solid curve shows the total intensity, which is a sum of the two components.
B. Comparison of the normalized total intensity with the data. }
\label{add25R}
\end{figure}

\begin{figure}
\noindent\includegraphics[scale=0.8]{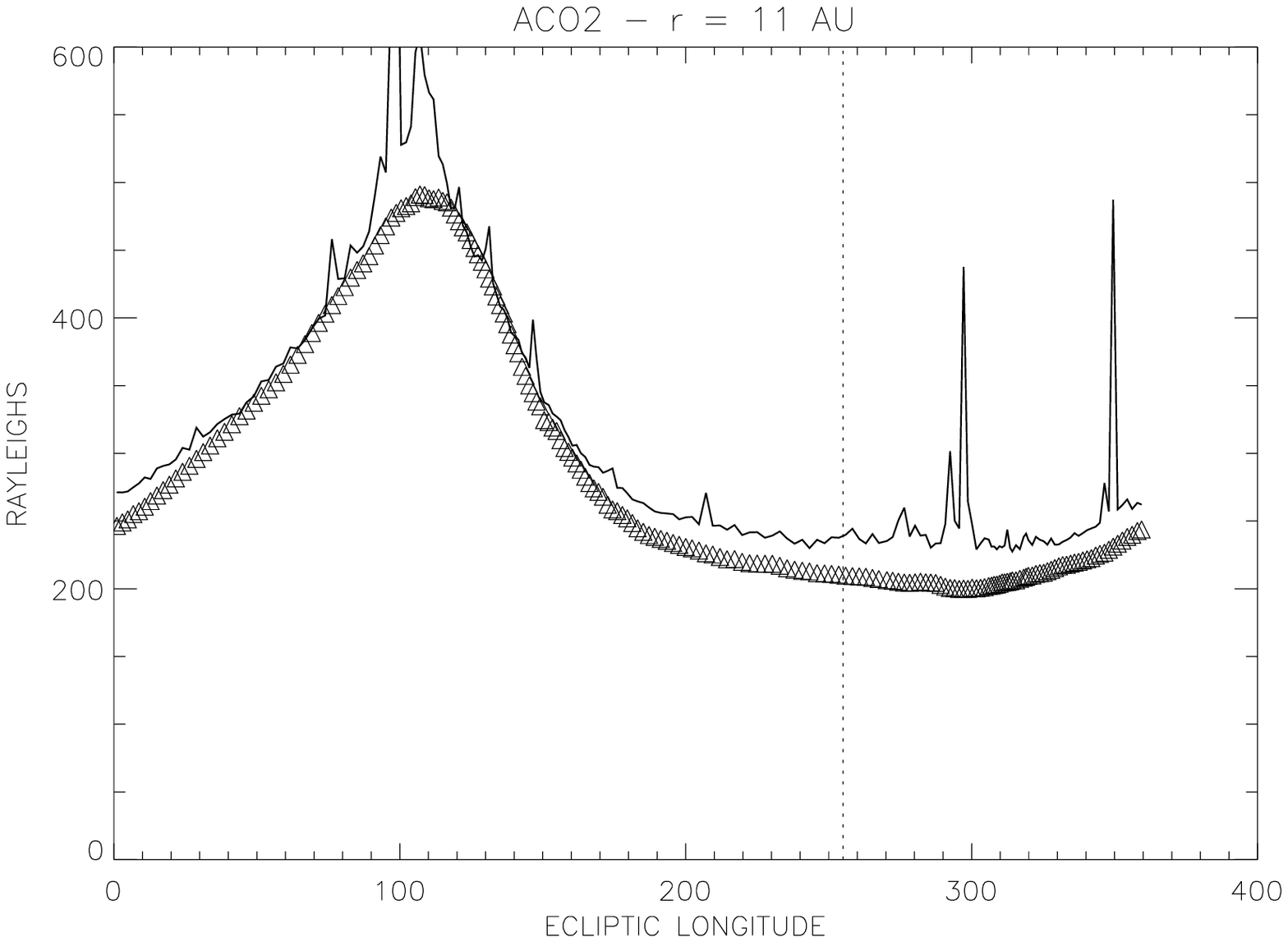}
\caption{Data and model comparison for the Alice Observations of the ACO-2 Roll performed by New Horizons at 11.3 AU from the Sun.
The X-axis shows the ecliptic longitude and the Y-axis is in Rayleigh. The spikes correspond to stars crossing the field of view of Alice.
The data show an excess over the model in the upwind hemisphere of about 20 R, very similar to the ACO-4 roll.}
\label{aco2}
\end{figure}

\begin{figure}
\noindent\includegraphics[scale=0.8]{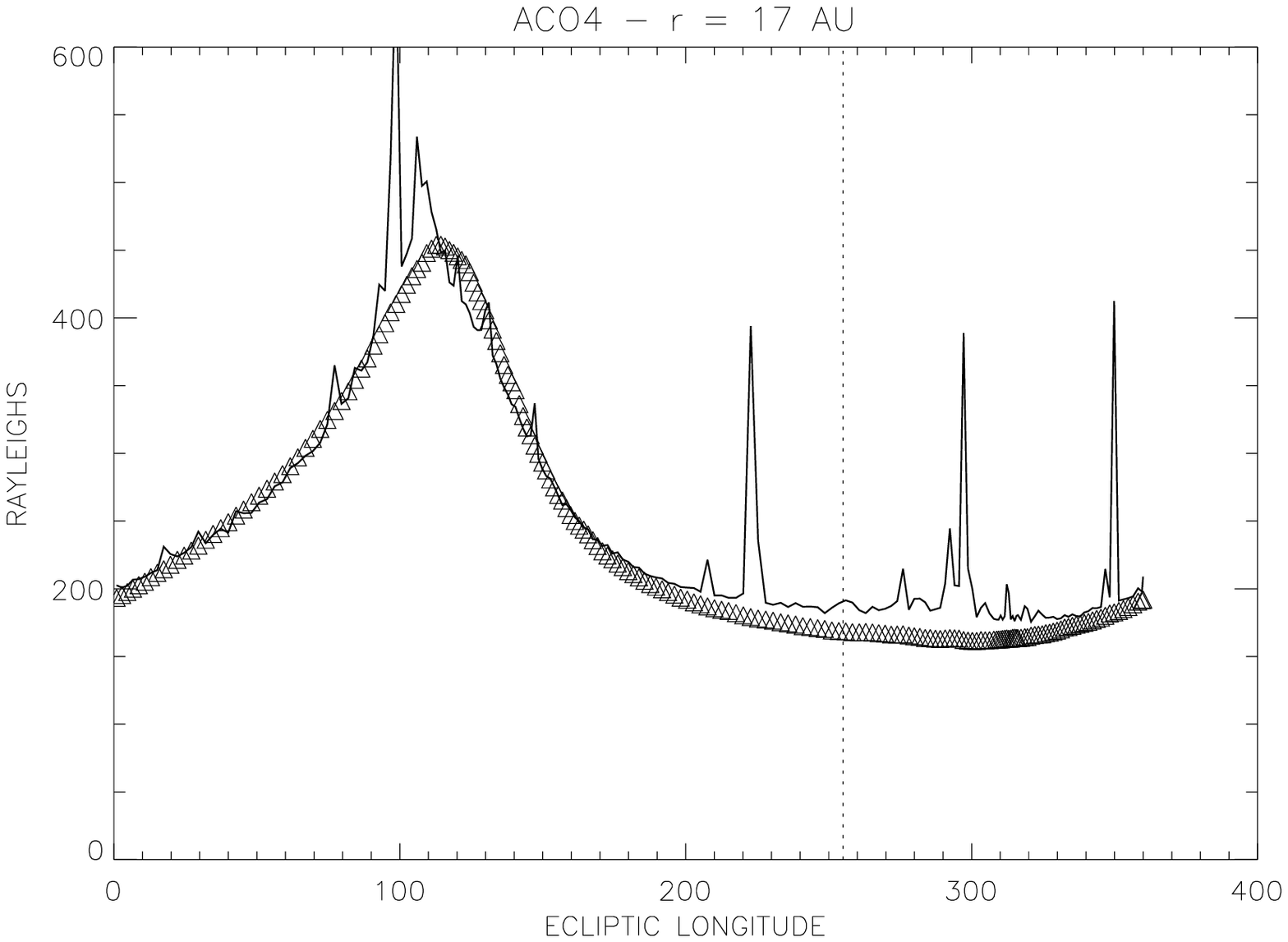}
\caption{Data and model comparison for the Alice Observations of the ACO-4 Roll performed by New Horizons at 17.0 AU from the Sun.
The X-axis shows the ecliptic longitude and the Y-axis is in Rayleigh. The spikes correspond to stars crossing the field of view of Alice.
The data show an excess over the model in the upwind hemisphere of about 20 R, very similar to the ACO-2 roll.}
\label{aco4}
\end{figure}

\begin{figure}
\noindent\includegraphics[scale=0.5]{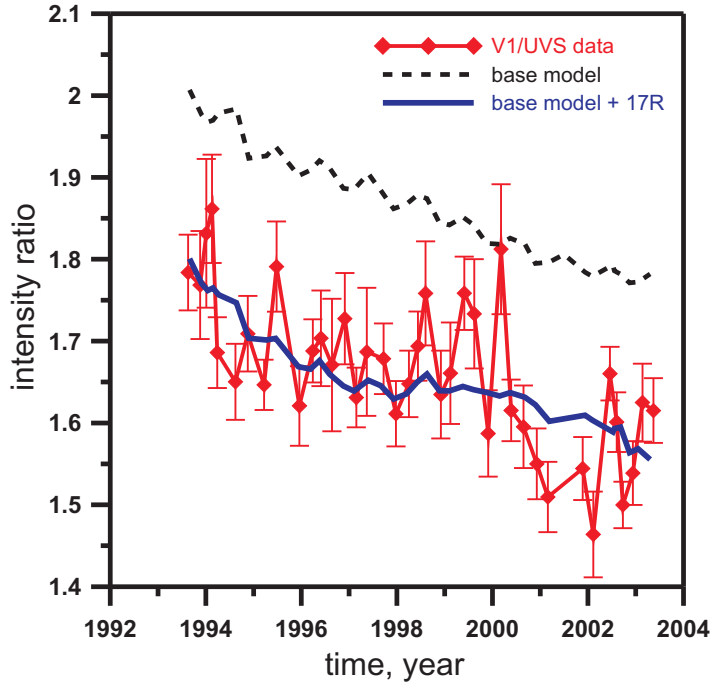}
\caption{Ratio of the Lyman~$\alpha$ intensities measured by Voyager~1 in 1993-2003 in the tail and nose directions.
Symbols show the data with error bars, the dashed curve corresponds to the results of the base model, and the solid curve corresponds
to the results of the base model with an artificially added 17~R of uniform extra-heliospheric emission.}
\label{ratio}
\end{figure}

\begin{figure}
\noindent\includegraphics[scale=0.5]{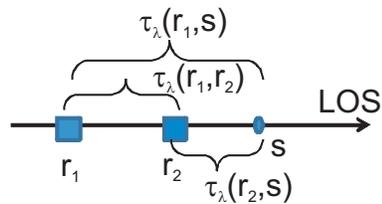}
\caption{Schematic representation of the optical depth between two observer's positions ($r_a$ and $r_2$) and scattered point ($s$).
Here it is assumed that the observer moves along its line-of-sight (LOS).}
\label{tau}
\end{figure}

\end{document}